\documentclass[compsoc,conference,a4paper,10pt]{IEEEtran}
\IEEEoverridecommandlockouts
\usepackage{cite}
\usepackage{amsmath}
\usepackage{amssymb}
\usepackage{amsfonts}
\usepackage{algorithmic}
\usepackage{graphicx}
\usepackage{textcomp}
\usepackage{bmpsize}
\usepackage{xcolor}
\usepackage[colorlinks=true,urlcolor=black]{hyperref}
\def\BibTeX{{\rm B\kern-.05em{\sc i\kern-.025em b}\kern-.08em
    T\kern-.1667em\lower.7ex\hbox{E}\kern-.125emX}}

\usepackage{pifont}
\usepackage{seqsplit}
\usepackage{booktabs}
\usepackage{multirow}
\usepackage{circledsteps}
\usepackage{array}
\usepackage{xurl}

\newcommand{\cmd}[1]{{\texttt{\seqsplit{#1}}}}

\newcommand{\cmark}{\ding{51}}
\newcommand{\xmark}{\ding{55}}

\def\aname{CTRAPS}
\def\toolkit{\texttt{\aname}}

\def\attacks{eleven}

\def\aone{Delete discoverable credentials}

\def\atwo{Factory reset authenticator}

\def\athree{Track user from credentials}

\def\afour{Fill authenticator's credential storage}

\def\afive{Force authenticator lockout}

\def\asix{Authenticator DoS}

\def\aseven{Profile authenticator}

\def\vulns{eight}
\def\vone{Unauthenticated CTAP client}
\def\vtwo{No authenticator feedback about API calls}
\def\vthree{NFC range provides \emph{UP}}
\def\vfour{Weak access control to destructive APIs}
\def\vfive{User trackable via CredId and UserId}
\def\vsix{\texttt{Reset} does not verify the user}
\def\vseven{\texttt{CredMgmt} allows to delete multiple credentials at once}
\def\veight{\texttt{Selection} is usable for DoS}

\def\apis{seven}

\def\auths{six}

\def\discrps{eight}

\def\ndiscrps{two}

\def\rps{ten}

\def\trans{two}

\def\counters{eight}
\def\cone{Trusted CTAP clients}
\def\ctwo{Authenticator visual feedback}
\def\cthree{User interaction for \emph{UP} over NFC}
\def\cfour{Dedicated PIN for destructive APIs}
\def\cfive{Dynamic and \emph{UV}-protected CredId and UserId}
\def\csix{\texttt{Reset} must require \emph{UV}}
\def\cseven{\texttt{CredMgmt} must require \emph{UP}}
\def\ceight{Rate limiting \texttt{Selection} calls}

\begin{document}

\title{\aname: CTAP Client Impersonation and API Confusion on FIDO2}

\author{\IEEEauthorblockN{Marco Casagrande}
\IEEEauthorblockA{\textit{Department of Digital Security} \\
\textit{EURECOM}\\
Sophia Antipolis, France \\
marco.casagrande@eurecom.fr}
\and
\IEEEauthorblockN{Daniele Antonioli}
\IEEEauthorblockA{\textit{Department of Digital Security} \\
\textit{EURECOM}\\
Sophia Antipolis, France \\
daniele.antonioli@eurecom.fr}
}

\maketitle

\begin{abstract}

FIDO2 is a popular technology for single-factor
and second-factor authentication. It is specified in an open
standard including the WebAuthn and CTAP application layer protocols. We focus
on CTAP which allows the communication between FIDO2 clients
and authenticators. No prior work explored
the CTAP Authenticator API which is a critical protocol-level attack surface as it deals with
credential creation, deletion, and management.
We address this gap by presenting the first security and privacy evaluation of the CTAP Authenticator API.
We uncover two classes of CTAP protocol-level attacks we call \aname.

The client impersonation (CI) attacks exploit the lack of client authentication to tamper with
FIDO2 authenticators. They include zero-click attacks capable of deleting
FIDO2 credentials, including passkeys, without user interaction.
The API confusion (AC) attacks abuse the lack of protocol API enforcements and confound
FIDO2 authenticators, clients, and users into calling unwanted CTAP
APIs while thinking they are calling legitimate ones. For example, a victim thinks is
authenticating to a website, when they are deleting their credentials.
The CTRAPS attacks are conducted either in proximity or remotely and
are effective regardless of the underlying CTAP transport (USB, NFC, or BLE).

We detail the \vulns\ vulnerabilities in the CTAP specification enabling the
\aname\ attacks. Seven of them are novel and include unauthenticated CTAP
clients and trackable FIDO2 credentials.
We release \toolkit, an original toolkit to analyze CTAP
and conduct the \aname\ attacks.
We confirm the attacks' feasibility by exploiting \auths\
popular authenticators, including a FIPS-certified one, from Yubico, Feitian,
SoloKeys, and Google, and \rps\ widely used relying parties,
such as Microsoft, Apple, GitHub, and Facebook.
We discuss \counters\ backward-compliant countermeasures
to fix the attacks and their root causes.
We responsibly disclosed our findings to the FIDO alliance and the affected vendors.

\end{abstract}

\begin{IEEEkeywords}
FIDO, CTAP, API Confusion
\end{IEEEkeywords}

\section{Introduction}\label{sec:intro}

\emph{Fast IDentity Online v2 (FIDO2)} is the de-facto standard
for single-factor (passwordless) and second-factor (2FA)
authentication. Google, Dropbox, and GitHub~\cite{lang2016security} designed
FIDO to offer a practical and scalable solution for authentication. FIDO has
been widely adopted  by industries and organizations including
Apple, Microsoft, and the US government~\cite{sales-usgov}.
Market forecasts predict the FIDO market to rapidly grow
from USD 230.6 million in 2022 to USD 598.6 million in 2031~\cite{sales-forecast}.
Yubico, a FIDO authenticator market leader, sold more than 22 million YubiKey
authenticators~\cite{yubico-sold}. This growth will continue because of the recent
industry-wide push towards single-factor passkey-based
authentication~\cite{passkeys-push,gh-2fa,google-passkeys}.

FIDO2 involves three entities: an \emph{authenticator}
that generates and asserts possession of authentication credentials (e.g., public-private key pairs),
a \emph{relying party} that authenticates the user (e.g., challenge-response protocol based on credentials),
and a \emph{client} who wants to authenticate to the relying party and manages the communication
between the authenticator and the relying party.
Typically, the authenticator is a dongle, the relying party is a web server,
and the client is a web browser or a mobile app.

The authenticator and the client communicate using the \emph{Client to Authenticator Protocol (CTAP)}.
CTAP works at the application-layer and is transported over
Universal Serial Bus (USB), Near Field Communication (NFC), or Bluetooth Low
Energy (BLE). It exposes the client to the \emph{CTAP Authenticator API}, usable
to interact with the authenticator, e.g., credential creation, management, and deletion.
These API calls might require User Verification (\emph{UV}) and User Presence
(\emph{UP}) authorization.

This work focuses on the CTAP protocol and its security and privacy
guarantees. There are only a few research studies about CTAP.
The authors of~\cite{barbosa2021provable} performed a provable security analysis on CTAP,
highlighting unauthenticated DH key exchange.
In a follow-up work~\cite{barbosa2023rogue}, they proposed an impersonation attack
exploiting CTAP to register an authenticator with an arbitrary relying party.
The authors in~\cite{guan2022formal} present a Machine-in-the-Middle (MitM) attack on CTAP
resulting in a privacy leak.
Other works target the authenticator with fault injection and side channel
attacks~\cite{titanninjaphysical,roche2021side}.

No prior work investigated the FIDO2 \emph{CTAP Authenticator API}. This API is a
critical protocol-level attack surface as it enables the creation,
management, and deletion of credentials and the administration of
authenticators. FIDO2 credentials are security
and privacy critical as they authorize access to popular online services,
including, social media, banking, data sharing, and e-commerce. A
protocol-level attack on the CTAP Authenticator API would enable access to
and manipulation of any credential stored on any authenticator,
regardless of the authenticator's hardware and software details. Hence, it is
crucial to assess the API's expected security and privacy properties and if they
hold in practice.

We fill this gap by presenting the first security and privacy assessment of
the CTAP Authenticator APIs. We uncover two attack classes
and eleven related attacks on CTAP that we call \textbf{\aname}.
The \emph{client impersonation (CI)} attacks exploit
the lack of client authentication to tamper with an authenticator.
Among others, they allow factory resetting an authenticator without
user interaction.
The \emph{API confusion (AC)} attacks abuse the lack of protocol API
enforcements and confound a FIDO2 authenticator, a client, and a user
into calling unwanted CTAP Authenticator APIs while believing they are
calling legitimate ones. For instance, a user thinks to be authenticating to a website but
they are instead deleting their authenticator credentials.

We consider two attacker models:
a CI attacker impersonating a CTAP client
and an AC attacker with a MitM position between the client and the authenticator.
The adversaries perform the attacks in \emph{proximity} or \emph{remotely}.
They do not require physical access to the authenticator, e.g., no side
channel or fault injection. Moreover, they do not need to compromise the
client or the authenticator, e.g., no client or authenticator malware.

The \aname\ attacks have a \emph{critical} and \emph{widespread} impact on the FIDO2 ecosystem.
They are critical as they violate the security, privacy, and availability of FIDO2 devices.
For example, a CI or an AC attacker can factory reset an authenticator deleting
all FIDO2 credentials and locking out the victim from the related service.
Despite targeting CTAP, the attacks also impact FIDO2 relying parties.
For example, they invalidate the non-discoverable credentials stored by the
relying party.
They are widespread as they exploit protocol-level vulnerabilities in the CTAP
application-layer protocol. Hence, they can be conducted against any FIDO2
device regardless of whether CTAP is transported over USB, NFC, or BLE.

We isolate \emph{\vulns\ vulnerabilities} in the CTAP specification enabling the \aname\ attacks.
Seven of them are novel within FIDO2.
They include unauthenticated CTAP clients, trackable credentials,
and weak authorization of (destructive) API calls.
The vulnerabilities are \emph{severe} as they affect authenticators
and clients implementing CTAP v2.0, v2.1, and v2.2.
We also find and disclose an implementation flaw on Yubico's authenticator firmware,
allowing an attacker to leak sensitive data and track users.
Yubico addressed the problem and assigned it CVE-2024-35311~\cite{yubico-cve}.

We present \toolkit, a new toolkit to experiment with CTAP and conduct the CTRAPS attacks.
The toolkit has three modules: CTAP testbed, CTAP clients, and Wireshark dissectors.
The testbed provides virtual clients and relying parties,
enabling local testing of the attacks without the involvement of actual devices.
The CTAP clients module performs the CI and AC attacks.
We implemented them to work from proximity and remotely.
Our CTAP clients allow the testing of the attacks on real-world authenticators and clients.
For example, we release an Android app and a Proxmark3 script to test the CI attacks over NFC.
The dissectors module includes an enhanced FIDO2 dissector for Wireshark
praising new and useful packet information
such as status codes and support for credential management.

We evaluate popular FIDO2 authenticators, clients, and relying parties.
We deploy them from proximity and remotely, testing different CTAP transports (USB and NFC).
We attack \emph{\auths\ authenticators} from Yubico, Feitian, SoloKeys, and Google.
One authenticator from Yubico is FIPS-compliant,
meaning that it utilizes cryptographic algorithms guaranteeing strict security standards.
We also exploit \emph{\rps\ relying parties} offering passkeys
and second-factor authentication, including Microsoft, Apple, GitHub, and Facebook.

We discuss \counters\ backward-compliant countermeasures
that fix the \aname\ attacks and their root causes.
The fixes include CTAP client authentication, stricter authorization requirements for destructive APIs,
introduce a dedicated PIN for destructive operations (e.g., credential deletion),
and rotate user identifiers and credentials to mitigate user tracking.
The countermeasures are backward-compliant as they rely on mechanisms already available in the
authenticator (e.g., PIN and LED) and do not require extra hardware (e.g., adding a display).

We summarize our contributions as follows:
\begin{itemize}
    \item We perform the first assessment of the CTAP Authenticator API. We
    unveil two classes of CTAP protocol-level attacks: CI and AC. The attacks
    compromise the security, privacy, and availability of the FIDO2 ecosystem.
    For instance, they (remotely) delete FIDO2 credentials,
    track users via FIDO2 credentials, and DoS authenticators.
    They are enabled by \vulns\ CTAP protocol level vulnerabilities,
    seven of which are new.
    \item We provide a toolkit to evaluate the CTAP Authenticator API
    and test our attacks in a virtual environment and on actual devices.
    We successfully conduct our attacks against
    \auths\ authenticators, \trans\ transports, and \rps\ relying parties.
    \item We design \counters\ backward-compliant countermeasures
    to fix our attacks and their root causes. We also responsibly disclosed
    our findings to the FIDO2 Alliance and affected vendors.
\end{itemize}

\textbf{Responsible disclosure}.
We responsibly disclosed our findings to the FIDO Alliance in November
2023~\cite{fidoall-securitysecr}. They acknowledged our report and shared it
with their members.
In May 2024, they provided feedback highlighting that the CI and AC attacks deployed by a
proximity-based attacker are less scalable than remote ones.
They argued on the effectiveness of CTRAPS attacks against authenticators running on a TEE.
They also discussed the possible addition of our attacks to FIDO's threat model.

In December 2023, we reported our findings to the affected
authenticator manufacturers (i.e., Yubico, Feitian, SoloKeys, and Google).
Google confirmed our findings, assigning them priority P2 and severity S2.
They responded that our attacks required a compromised FIDO client
and closed the issue without resolution. We argue that Google's
assessment is incorrect as our attacks do not require a compromised FIDO client.
Yubico confirmed the implementation bug we found, pushed a fix in production,
published a security advisory~\cite{yubico-advisory}, and assigned it CVE-2024-35311~\cite{yubico-cve}.
The other manufacturers acknowledged the report without commenting on it.

We also contacted Apple and Microsoft regarding their weak credential
protection policy that facilitates user tracking and profiling.
They responded that our report has no security implications for their products.

\textbf{Ethics and availability}.
We conducted our experiments ethically. We evaluated our 
authenticators and accounts. We did not collect personal data and involved
third parties. To advance open science, we open source our contributions,
including the \texttt{CTRAPS} toolkit, found at \url{https://github.com/Skiti/CTrAPs}.

\section{Background and System Model}\label{sec:back}

We introduce FIDO2, CTAP, and our system model.

\subsection{FIDO2}\label{subsec:back-fido}

FIDO2~\cite{fidoall-specs} is an open and pervasive standard for single-factor
and multi-factor authentication. It is managed by the FIDO Alliance.
FIDO2 has four entities: an authenticator, a client, a user, and a relying party.
In a typical scenario, a user connects their authenticator to the client
to access an online service hosted by a relying party.

The FIDO2 specification includes the WebAuthn and CTAP application-layer protocols.
WebAuthn provides a secure communication channel to a
relying party and a client. Its latest version is WebAuthnL2~\cite{webauthn2-standard}.
CTAP, the focus of this work, enables a secure connection
between a FIDO2 authenticator and a client via the CTAP Authenticator API.
For example, the \cmd{MakeCred} API registers a new credential
while the \cmd{GetAssertion} API authenticates a credential.

A FIDO2 \emph{credential} is a key pair used to sign and verify authentication
challenges to authenticate a user. The digital signature is computed using
standard techniques, like Elliptic Curve Digital Signature Algorithm (ECDSA).
Access to the credential private key is guarded by encryption
using a credential master key, which is unique to each authenticator
and stored in the authenticator.

FIDO2 credentials can be \emph{discoverable} or \emph{non-discoverable}.
Discoverable credentials, also known as passkeys, are stored on the
authenticator and used for passwordless authentication.
Non-discoverable credentials are stored by the relying party
and used for multi-factor authentication.

FIDO2 credentials are associated with a credential identifier (CredId),
a relying party identifier (RpId), and a user identifier (UserId).
The CredId uniquely identifies a FIDO2 credential
and is derived from the credential master key stored in the authenticator.
When FIDO2 clients authenticate a credential, they must know its associated CredId.
The RpId indicates the relying party with which the credential was registered.
It is public as it corresponds to the base domain of the relying party (e.g., \texttt{login.microsoft.com}).

The UserId represents the user's online account within the relying party's service.
The relying party assigns a random UserId to the user during account registration,
and it is shared across all FIDO credentials associated with that user.
The optional FIDO2 \emph{CredBlob} extension allows a relying party
to store additional metadata inside a credential.

\subsection{CTAP}\label{subsec:back-ctap}

The Client-to-Authenticator Protocol (CTAP) is a core part of the FIDO2 standard,
alongside WebAuthn.
It is an application-layer protocol that defines the communication
between a FIDO client and an authenticator.
CTAP has considerably evolved since its inception.
CTAP1, also known as FIDO U2F (Universal 2nd Factor), introduced a second-factor
authentication mechanism to combat phishing.
CTAP2.0 maintains backward compatibility with CTAP1 while introducing
passwordless (single-factor) authentication.
CTAP2.1~\cite{ctap21-standard} adds the credential protection policy,
discoverable credential management (i.e., the \cmd{CredMgmt} API), and biometric authentication.
CTAP2.2~\cite{ctap22-draft}, the latest CTAP version still considered a draft,
supports hybrid authenticators and QR codes.

CTAP relies on two user authorization mechanisms to secure API calls from the client:
(i) \emph{User Verification (UV)}, which requires the user to enter a PIN or biometric data,
and (ii) \emph{User Presence (UP)}, which requires the user to press a button on the authenticator
or to bring it into the client's NFC range.

Table~\ref{tab:apis-uvup} shows the seven \emph{CTAP Authenticator APIs} studied in
this paper and their \emph{UV} and \emph{UP} requirements:
\begin{itemize}
    \item [\cmd{MC}:] \cmd{MakeCred} registers a new credential bound to an online account with a relying party.
	\item [\cmd{GA}:] \cmd{GetAssertion} authenticates to a relying party by proving possession of a credential.
	\item [\cmd{CM}:] \cmd{CredMgmt} enumerates, modifies, and deletes the authenticator's discoverable credentials.
	\item [\cmd{CP}:] \cmd{ClientPin} handles \emph{UV} based on a user PIN to be submitted via the client's UI.
	\item [\cmd{Re}:] \cmd{Reset} wipes all discoverable and non-discoverable credentials and generates a new master key.
	\item [\cmd{Se}:] \cmd{Selection} selects an authenticator to operate among the available ones.
	\item [\cmd{GI}:] \cmd{GetInfo} returns the authenticator's details, like manufacturer, transports, extensions, and settings.
\end{itemize}

\begin{table}[tb]
  \caption{CTAP Authenticator APIs with their \emph{UV}
  and \emph{UP} authorization requirements, and support for subcommands.
  Yes$^1$: depends on the client and relying party configuration,
  Yes$^2$: depends on API subcommand.
}
  \renewcommand{\arraystretch}{1.1}
  \centering\small
  \begin{tabular}{@{}lccc@{}}
    \toprule
    \textbf{CTAP API}          & \textbf{UV} & \textbf{UP} & \textbf{Subcmd}\\
    \midrule
    \texttt{MakeCred} (\texttt{MC})  & Yes         & Yes  & No        \\
    \texttt{GetAssertion}  (\texttt{GA})   & \hspace{1.7mm}Yes$^1$     & \hspace{1.7mm}Yes$^1$  & Yes    \\
    \texttt{CredMgmt}      (\texttt{CM})  & Yes     & No & Yes           \\
    \texttt{ClientPin}          (\texttt{CP})   & \hspace{1.7mm}Yes$^2$     & No & Yes          \\
    \texttt{Reset}                (\texttt{Re})    & No          & Yes & No        \\
    \texttt{Selection}         (\texttt{Se})    & No          & Yes & No        \\
    \texttt{GetInfo}          (\texttt{GI})     & No          & No & No         \\
    \bottomrule
  \end{tabular}
\label{tab:apis-uvup}
\end{table}

The \cmd{GetAssertion}, \cmd{CredMgmt}, and \cmd{ClientPin} APIs
have API subcommands.
For example, \cmd{CredMgmt(GetCredsData)} returns the number of stored discoverable
credentials and \cmd{CredMgmt (DelCreds)} deletes all discoverable credentials.
Some API subcommands, compared to their original API, have more relaxed requirements.
For instance, \cmd{ClientPin(KeyAgreement)} does not require \emph{UV}.

CTAP offers other optional security and privacy mechanisms.
The authorization requirements for \cmd{GetAssertion} depend on the client and relying party configuration.
A client can specify the option \emph{up=false} to skip \emph{UP}.
At registration time, a relying party can enforce access control
by specifying a credential protection policy via the optional \emph{CredProtect} extension.
However, the default policy skips \emph{UV}, resulting in weak privacy protection.

\subsection{System Model}\label{subsec:sm}

\begin{figure}[tb]
  \centering
  \includegraphics[width=1\linewidth]{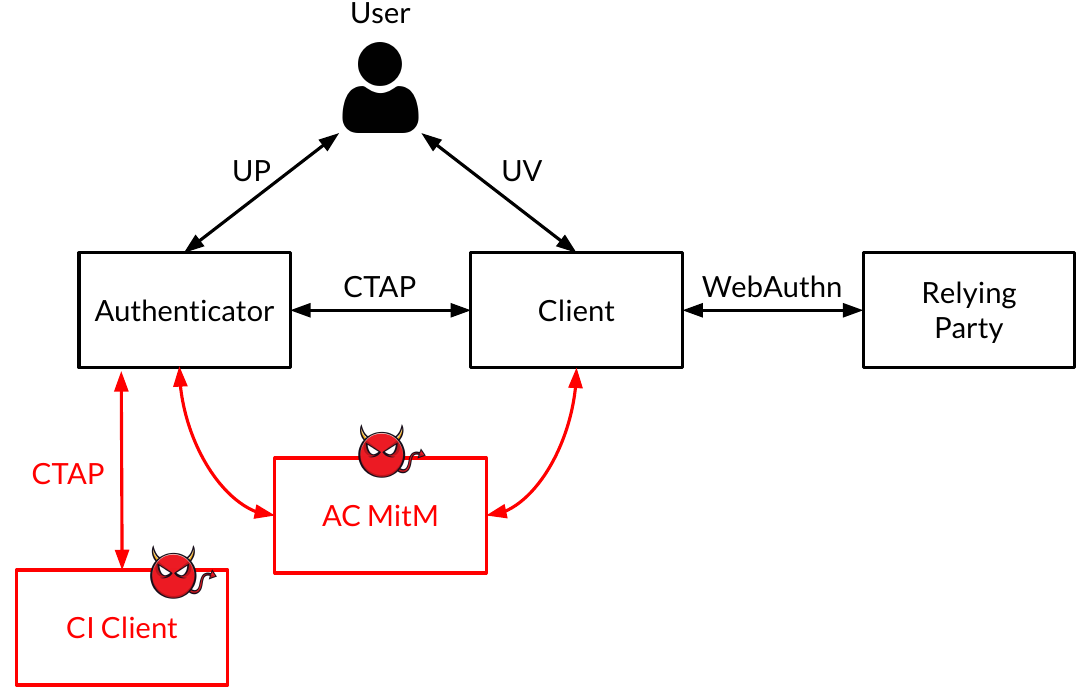}
  \caption{\textbf{\aname\ threat model.} The user authenticates to the relying party
  using a client (e.g., browser) and an authenticator (hardware dongle). The
  user when needed grants UP by pressing a button on to the authenticator
  and UV by submitting a PIN to the client.
  We study two attacker models: (i) a client impersonation attacker
  targeting the authenticator over CTAP (left), (ii) a MitM attacker in the
  CTAP channel between the authenticator and the client.}
  \label{fig:threat-model}
\end{figure}

We adopt the official FIDO2 system model~\cite{fidoall-specs}.
Figure~\ref{fig:threat-model} shows the system model's four entities:
authenticator, client, relying party, and user.
The user connects the authenticator to the client to authenticate on a service
hosted by the relying party. The entities support up to CTAP2.2 and WebAuthnL2 (i.e., the latest
and supposedly most secure FIDO2 protocol versions).
Next, we describe each entity. The attacker models are presented in
Sections~\ref{subsec:ci-am} and~\ref{subsec:ac-am} as they are specific to the CI and AC attacks.

\textbf{Authenticator}.
The authenticator is a FIDO2 authenticator: a user device
that can be connected to the client (e.g., a USB/NFC dongle).
The authenticator runs a CTAP server that exposes the CTAP Authenticator API.
The API is accessible over USB, NFC, and BLE.
The authenticator supports FIDO2's \emph{UP}
and \emph{UV} user authorization mechanisms.
It stores discoverable credentials and the credential master key.

\textbf{Client}.
The client is a FIDO2 client handling the communication between the
authenticator and the relying party. It exposes a CTAP client to the
authenticator and a WebAuthn client to the relying party. The client could be
a web browser, a mobile app for Android~\cite{app-feitian-android} or iOS~\cite{app-feitian-ios},
or a command line tool like the Yubico CLI~\cite{cli-yubico}.

\textbf{Relying party}.
The relying party is an online service that relies on FIDO2 passwordless or
multi-factor authentication. It runs a WebAuthn server that responds to FIDO2 registration
and authentication requests. The relying party stores
\seqsplit{non-discoverable} credentials, and user and credential identifiers. 
The relying party communicates with the client using TLS.
Offline operations on the authenticator, like deleting discoverable credentials,
indirectly affect the relying party by making the user unable to log into their online service.

\textbf{User}.
The user owns an authenticator
and a device that runs the FIDO2 client, e.g., a YubiKey dongle and a laptop.
They utilize their authenticator to register FIDO2 credentials
and authenticate to the associated relying party.
To do so, they connect their authenticator to the client
and provide \emph{UV} and \emph{UP}, if necessary.
The user manages the authenticator via the client,
without connecting to a relying party.
For example, they can check their discoverable credentials
and change the authenticator's PIN.

\section{CTRAPS Client Impersonation Attacks}\label{sec:ci}

The \emph{CTRAPS CI attacks} target an authenticator while spoofing a client
to perform CTAP API calls without user authorization.
CI attacks factory reset the authenticator via the \cmd{Reset} API,
track the user via \cmd{GetAssertion},
lock the authenticator via \cmd{ClientPin},
and profile the authenticator via \cmd{GetInfo}.
They exploit five protocol-level CTAP vulnerabilities we found (described in Section~\ref{subsec:vulns}).
For instance, the absence of CTAP client authentication facilitates impersonation,
the use of NFC transport allows to bypass \emph{UP},
and the lack of \emph{UV} when calling \cmd{Reset} enables unauthorized factory resets.

The attacks advance the state of the art in FIDO2's security and privacy by introducing client impersonation.
This is a new class of attacks previously unseen in FIDO2, as shown in Table~\ref{tab:rw-compare}.
The CI attacks require limited or no user interaction, depending on the CTAP transport.
For example, by using NFC, they bypass \emph{UP}, leading to zero-click attacks.
The CI attacks also involve \emph{no client compromise},
being deployed from a client owned by the attacker.
Next, we introduce the CI attacker model and describe the attacks.

\subsection{CI Attacker Model}\label{subsec:ci-am}

The CI attacker model assumes an attacker
impersonating a CTAP client to the victim's authenticator,
referenced as \textcolor{red}{CI Client} in Figure~\ref{fig:threat-model}.
The attacker is in proximity of the victim's authenticator, or can remotely connect to it.
They have no physical access to and do not tamper with the victim's client and authenticator.
They do \emph{not} install malware on the victim's device running the FIDO2 client.

The CI attacker model maps to several relevant attack scenarios.
For example, they can approach a target authenticator over NFC while impersonating
a client (e.g., via a smartphone or a Proxmark), place a malicious NFC device in a place
where a user might touch it with an authenticator (e.g., under a table),
They can also communicate with the victim's authenticator using a compromised
hardware device, such as a USB hub that connects the user's machine and the authenticator,
or virtual USB peripheral, through a setup similar to~\cite{virtualfido-ci}.

\subsection{CI Attacks Description}\label{subsec:ci-att}

We describe the four CI attacks, which we label CI$_1$, CI$_2$, CI$_3$, and CI$_4$.

\begin{figure}[tb]
    \centering
    \includegraphics[width=0.8\linewidth]{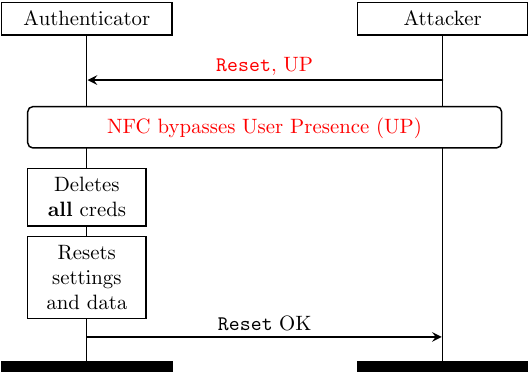}
    \caption{\textbf{CI$_{1}$ attack.} \atwo\ via \texttt{Reset}.
    While in NFC range, the attacker calls the \texttt{Reset} API.
    Over NFC, the authenticator skips \emph{UP} and instantly factory resets,
    deleting all of its discoverable and non-discoverable credentials.}
    \label{fig:ci1-pb-reset}
\end{figure}

\textbf{CI$_{1}$: \atwo}.
In CI$_{1}$, the attacker abuses the \textcolor{red}{\cmd{Reset}} API
to factory reset an authenticator,
as shown in Figure~\ref{fig:ci1-pb-reset}.
The attacker connects to the authenticator and, without authenticating,
issues a factory reset command (which requires \emph{UP}).
Over USB, the attack requires one click (\emph{UP})
and the authenticator having been plugged into the USB port within the last ten seconds.
Over NFC, the attacker achieves zero-click reset
by exploiting a CTAP quirk intended to enhance usability.
That is, NFC communication inherently implies user presence,
allowing \emph{UP} to be bypassed.
The execution of the factory reset wipes out all credentials,
even the non-discoverable ones stored by the relying party,
as it erases the credential master key necessary for decryption. It also deletes
the authenticator's settings, including the PIN, user preferences, and stored data.
Then, the authenticator confirms the successful reset.

\begin{figure}[tb]
    \centering
    \includegraphics[width=0.9\linewidth]{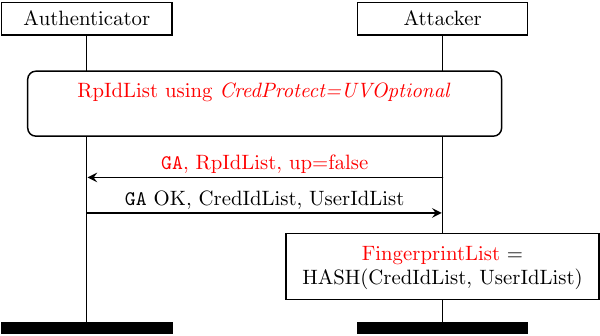}
    \caption{\textbf{CI$_{2}$ attack.} \athree\ via \texttt{GetAssertion}.
    The attacker connects to the authenticator and calls the \texttt{GetAssertion} API (\texttt{GA} in the figure).
    They skip \emph{UV} by targeting relying parties using the weak and default \emph{CredProtect} default policy
    and skip \emph{UP} by passing \emph{up=false}. The authenticator returns a list of credential and user identifiers,
    used by the attacker to fingerprint the authenticator and track the user.}
    \label{fig:ci2-rm-track}
\end{figure}

\textbf{CI$_{2}$: \athree}.
In CI$_{2}$, instead of using \textcolor{red}{\cmd{GetAssertion}} for authentication,
the attacker exploits it to leak identifying data and track the user,
as shown in Figure~\ref{fig:ci2-rm-track}.
CI$_{2}$ requires a pre-determined list of RpId
for which the attacker aims to leak credentials.
This is straightforward, as this information is publicly accessible.
Although the \cmd{GetAssertion} API requires both \emph{UV} and \emph{UP},
the attacker can circumvent both authorizations,
resulting in a \emph{zero-click} data leak and enabling user tracking.
They bypass \emph{UP} by issuing a \cmd{GetAssertion} command
containing the \emph{up=false} option.
They bypass \emph{UV} by only targeting relying parties that register credentials using
the weak and default \emph{CredProtect=UVOptional} policy, such as Microsoft and Apple.
Executing \cmd{GetAssertion} returns a list of credential and user identifiers.
These identifiers can be used to fingerprint the user
and to track them over multiple connections
by performing CI$_{2}$ each time and looking for matching fingerprints.
CI$_{2}$ also works on credentials protected
by stronger policies (i.e., \emph{CredProtect=UVRequired}
and \emph{CredProtect=UVOptionalWithCredIDList}),
but requires \emph{UV} or knowledge of the credential identifiers.

\textbf{CI$_{3}$: \afive}.
In CI$_{3}$, the attacker abuses the \textcolor{red}{\cmd{ClientPin}} API,
protecting the authenticator from PIN brute-forcing,
to lock the authenticator or even force a factory reset.
They submit to the authenticator several wrong PIN guesses in a row
via the \textcolor{red}{\cmd{ClientPin(GetPinToken)}} subcommand.
After three wrong guesses, the authenticator enters a soft lock mode
preventing actions until a reboot (i.e., leaving and re-entering
a client's NFC range, or detaching and re-attaching to a USB port).
After a maximum of failed PIN attempts (CTAP mandates eight),
the authenticator enters a hard lock mode only restorable through a factory reset,
which wipes out all credentials and can lead to account loss.

\textbf{CI$_{4}$: \aseven}.
In CI$_{4}$, the attacker calls \textcolor{red}{\cmd{GetInfo}}
to leak the authenticator's technical details.
This attack can be used as a stepping stone
to more advanced attacks, to profile the user and track them in future connections,
and to assess whether the authenticator is vulnerable
to an implementation-specific attack like~\cite{yubico-cve}.
The leaked details include the manufacturer, model, and FIDO2 version,
and the supported algorithms, transports, options, and extensions.
The authenticator also discloses user settings,
such as FIDO2 being disabled over a specific transport.

\section{CTRAPS API Confusion Attacks}\label{sec:ac}

The \emph{CTRAPS AC attacks} take advantage of a novel attack technique for FIDO,
which we refer to as \emph{API confusion}.
API confusion tricks a client, an authenticator, and their user
into calling a CTAP Authenticator API while they think they are calling a different one.
The called API has the same or lower \emph{UV} and \emph{UP} requirements of the intended API.
For example, AC attacks can erase FIDO2 credentials, including passkeys,
lock the user out of their authenticator, and track them.

AC is effective as it does not require social
engineering~\cite{ulqinaku2021real} or other deception
techniques~\cite{mahdad2024overlays} to trick the user into calling an unwanted API.
The user cannot detect an API confusion because
it requires expected  \emph{UV} or \emph{UP} actions.
The AC attacks exploit the eight protocol-level vulnerabilities we outline in Section~\ref{subsec:vulns}.
For instance, the absence of authenticator feedback during API calls grants stealthiness
and the use of static credential and user identifiers enables user tracking.

No prior work considered the AC attack vector for FIDO2, as shown in Table~\ref{tab:rw-compare}.
Existing attacks on FIDO include MitM on the Diffie-Hellman key exchange, CTAP traffic eavesdropping,
U2F impersonation, physical access, and side channel attacks on the authenticator.
Moreover, the AC attacks target the entire CTAP Authenticator API surface,
whereas previous research only focused on \cmd{ClientPin} and \cmd{MakeCred}.
Next, we will introduce the AC attacker model and attacks.

\subsection{AC Attacker Model}\label{subsec:ac-am}

The AC attacker model assumes a MitM attacker between the authenticator and the client,
referenced as \textcolor{red}{AC MitM} in Figure~\ref{fig:threat-model}.
The attacker is either in proximity to the authenticator and the client (e.g., an NFC skimmer)
or can contact them from remote (e.g., a remotely controllable USB hub).
They are unable to modify the authenticator's firmware
or compromise a legitimate FIDO2 client and relying party.
They have no physical access to the authenticator.

An AC attacker model has several associated real-world attack scenarios.
For example, they can get a MitM position over NFC interposing an NFC skimmer
between the client and the authenticator.
They can achieve a MitM position over USB
with setups such as those discussed in~\cite{rijnard2014usbmitm} and~\cite{badusb-mitm}.
For instance, the attacker can remotely compromise a USB device connected
to the user's device running the FIDO2 client,
such as a USB hub that routes traffic between other USB peripherals.

Alternatively, they can gain privileged access via techniques like UACMe~\cite{uacme-mitm}
and then leverage USBPcap to USB MitM a victim's Windows machine running the FIDO2 client.
The attacker can also install on the user's machine a malicious app
exploiting libraries that provide access to USB HID traffic.
We implement this attack scenario in Section~\ref{subsec:imp-reader-impers}
by developing an Electron app that mimics a MitM attacker using the \emph{node-hid} module.

\subsection{AC Technique and Combinations}\label{subsec:ac-sc}

\begin{table}[tb]
  \renewcommand{\arraystretch}{1.1}
  \caption{There are 49 ways to perform AC against 7 CTAP Authenticator APIs.
  The user intends to call \texttt{API\;A}, instead is tricked into
  calling \texttt{\color{red} API\;B}.
  \cmark$^1$: proximity-based attacker,
  \cmark$^2$: default \emph{CredProtect=UVOptional} if credential protection is enabled,
  n/a: not applicable.}
  \centering\small
  \begin{tabular}{@{}lccccccc@{}}
    \toprule
    & \texttt{\color{red} CM} & \texttt{\color{red} Re} & \texttt{\color{red} GA} & \texttt{\color{red} MC}
    & \texttt{\color{red} CP} & \texttt{\color{red} Se} & \texttt{\color{red} GI} \\
    \midrule
    \texttt{CM}  & n/a         & \hspace{1.5mm}\cmark$^1$  & \cmark\     & \hspace{1.5mm}\cmark$^1$  & \cmark\     & \cmark\     & \cmark\     \\
    \texttt{Re}   & n/a         & n/a         & \hspace{1.5mm}\cmark$^2$  & n/a         & \cmark\     & \cmark\     & \cmark\     \\
    \texttt{GA}   & \cmark\     & \cmark\     & n/a         & \cmark\     & \cmark\     & \cmark\     & \cmark\     \\
    \texttt{MC}  & \cmark\     & \cmark\     & \cmark\     & n/a         & \cmark\     & \cmark\     & \cmark\     \\
    \texttt{CP}   & \cmark\     & \hspace{1.5mm}\cmark$^1$  & \cmark\     & \hspace{1.5mm}\cmark$^1$  & n/a         & \cmark\     & \cmark\     \\
    \texttt{Se}    & n/a         & \cmark\     & \hspace{1.5mm}\cmark$^2$  & n/a         & \cmark\     & n/a         & \cmark\     \\
    \texttt{GI}     & n/a         & \hspace{1.5mm}\cmark$^1$  & \hspace{1.5mm}\cmark$^2$  & \cmark\     & \cmark\     & \cmark\     & n/a         \\
    \midrule
    Total               & 3           & 6           & 6           & 4           & 6           & 6           & 6           \\
    \bottomrule
  \end{tabular}
  \label{tab:ac-overview}
\end{table}

The seven AC attacks rely on the API confusion attack technique.
The attacker intercepts a call to \texttt{API\;A}
and changes (i.e., confounds) it to \textcolor{red}{\cmd{API\;B}}.
This action only requires that \textcolor{red}{\cmd{API\;B}}
has the same or lower \emph{UV} and \emph{UP} authorization
requirements than \texttt{API\;A}.
The AC technique has six steps:

\begin{enumerate}
    \item The user calls \texttt{API\;A} through the client. The API might require \emph{UV}
        and/or \emph{UP}.
    \item If required by \texttt{API\;A}, the attacker obtains \emph{UV}
by executing the CTAP PIN/UV authentication protocol v1 (via \cmd{ClientPin}).
The user inputs the PIN on the client, which encrypts it and submits it to the authenticator.
The authenticator responds with an encrypted User Verification Token (UVT),
that will be attached to any API call requiring \emph{UV}.
    \item The attacker calls \textcolor{red}{\cmd{API\;B}} rather than
        \texttt{API\;A} based on the AC combinations in Table~\ref{tab:ac-overview}.
    \item If required by \texttt{API\;A}, the attacker obtains \emph{UP}
from the user, unable to realize they are under attack.
The attacker can only obtain \emph{UP} once,
as multiple requests would alarm the user.
This step is bypassed whenever NFC proximity implies \emph{UP}.
    \item The authenticator executes \textcolor{red}{\cmd{API\;B}} and returns a success message.
    \item The attacker informs the victim via the CTAP client that \cmd{API\;A} was successfully executed.
\end{enumerate}

The AC strategy is effective on \emph{7} CTAP Authenticator APIS and provides
\emph{49} ways to confound the victim as shown in Table~\ref{tab:ac-overview}.
Multiple (\cmd{API\;A}, \textcolor{red}{\cmd{API\;B}}) pairs achieve the same goal.
The amount of available pairs depends on their \emph{UV} and \emph{UP} requirements
and, in the case of AC$_{3}$, also on the \emph{CredProtect} policy.
The first column lists \apis\ APIs the user intends to call (\texttt{API\;A}),
and the remaining columns represent the API called by the attacker (\textcolor{red}{\cmd{API\;B}}).
For instance, AC$_{1}$ is available whenever the user calls \cmd{MakeCred}, \cmd{GetAssertion},
or \texttt{ClientPin}, confounding the call to \texttt{\color{red} CredMgmt}.
Some combinations are only feasible by a proximity-based attacker
or under the default \emph{CredProtect} policy.
An API cannot be confounded with itself
or APIs with incompatible authorization requirements.

\subsection{AC Attacks Description}\label{subsec:ac-att}

We describe seven AC attacks labeled AC$_1$, AC$_2$, AC$_3$,
AC$_4$, AC$_5$, AC$_6$, and AC$_7$. AC$_1$ exploits all possible ways to
call \texttt{\color{red} CM}, AC$_2$ does this with \texttt{\color{red} Re}, and so on.

\begin{figure}[tb]
    \centering
    \includegraphics[width=1.0\linewidth]{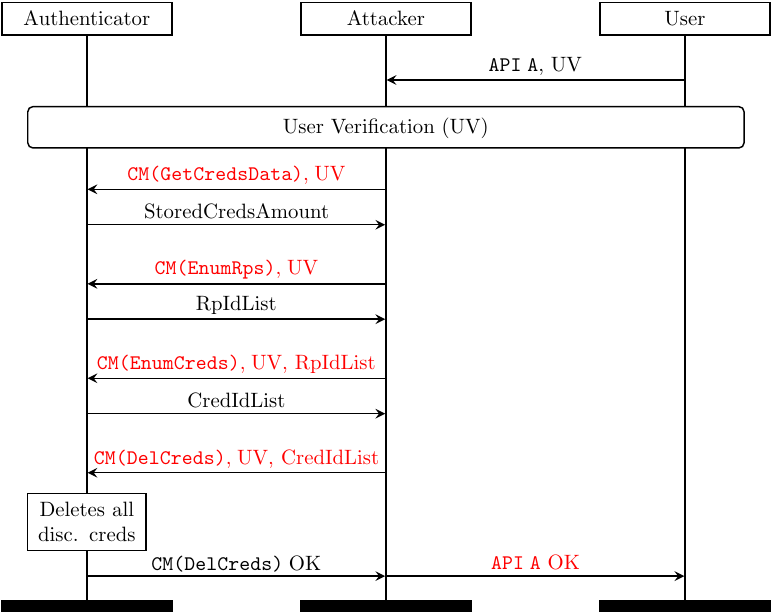}
    \caption{\textbf{AC$_{1}$ attack.} \aone\ attack with proximity.
    The user intends to call \texttt{API\;A}, requiring \emph{UV} but not necessarily \emph{UP}.
    For example, \texttt{GetAssertion}, \texttt{ClientPin}, or \texttt{MakeCred}.
    The attacker obtains \emph{UV} from the unsuspecting user.
    Instead of \texttt{API\;A}, they call \texttt{CredMgmt} (\texttt{CM} in the figure).
    They execute four \texttt{CredMgmt} subcommands which list
    and then delete all discoverable credentials on the authenticator.}
    \label{fig:ac1-pb-delete}
\end{figure}

\textbf{AC$_{1}$: \aone}.
In AC$_{1}$, the attacker abuses the \cmd{CredMgmt} API to delete
all discoverable credentials stored on the authenticator, as shown in Figure~\ref{fig:ac1-pb-delete}.
The user intends to call \cmd{API\;A}, which requires \emph{UV} but not necessarily \emph{UP},
such as \cmd{GetAssertion}, \texttt{ClientPin}, or \cmd{MakeCred}.
Instead, the attacker executes four separate \cmd{CredMgmt} subcommands,
none of which require \emph{UP}.
First, they check the existence of discoverable credentials to erase (StoredCredsAmount)
via \cmd{CredMgmt(GetCredsMetadata)}.
Second, they retrieve the list of relying parties stored on the authenticator (RpIdList)
via \cmd{CredMgmt(EnumRps)}.
Third, they use RpIdList to retrieve the list of stored credential identifiers (CredIdList)
via \cmd{CredMgmt(EnumCreds)}.
Fourth, they use CredIdList to delete all discoverable credentials
via \cmd{CredMgmt(DelCreds)}.
Finally, they falsely return \cmd{API\;A} OK to the user.

\textbf{AC$_{2}$: \atwo}.
In AC$_{2}$, the attacker exploits the \cmd{Reset} API to factory reset the authenticator, similar to CI$_{1}$.
Since \cmd{Reset} over USB requires \emph{UP}, but not \emph{UV}, an attacker can
confound \cmd{MakeCred}, \cmd{GetAssertion}, and \cmd{Selection} into a \cmd{Reset} call.
An attacker over NFC, able to bypass \emph{UP}, can also confound
\cmd{CredMgmt}, \cmd{ClientPin}, and \cmd{GetInfo}.

\textbf{AC$_{3}$: \athree}.
In AC$_{3}$, the attacker misuses the \cmd{GetAssertion} API to leak unique identifiers
as fingerprints and track the user, similar to CI$_{2}$.
They can confound \cmd{MakeCred}, \emph{CredMgmt},
and \emph{ClientPin} into a \cmd{GetAssertion} call,
if they want to access credentials protected by the \emph{\seqsplit{CredProtect=UVRequired}}
or \emph{\seqsplit{CredProtect=UVOptionalWithCredIDList}} policies.
Additionally, the attacker can also confound \cmd{Reset}, \cmd{Selection}, and \cmd{GetInfo}
if they only wants to access credentials protected
by the weak \emph{\seqsplit{CredProtect=UVOptional}} default policy.

\textbf{AC$_{4}$: \afour}.
In AC$_{4}$, the attacker repeatedly calls \cmd{MakeCred} to register new discoverable credentials,
until the authenticator's credential storage is full.
They exploit the \emph{rk=true} option to enforce the generation
of discoverable credentials over non-discoverable ones.
A filled storage compromises the authenticator's availability
as the user cannot register new discoverable credentials.

\textbf{AC$_{5}$: \afive}.
In AC$_{5}$, the attacker abuses the \cmd{ClientPin} API to lock the authenticator and force
a mandatory factory reset, similar to CI$_{3}$.
Although \cmd{ClientPin} requires \emph{UV},
the attacker wants to fail multiple PIN attempts (i.e., they do not need \emph{UV}).
Consequently, they can confound any API call into a \cmd{ClientPin} call,
as they do not need authorization.

\textbf{AC$_{6}$: \asix}.
In AC$_{6}$, the attacker calls \cmd{Selection} to trigger
an unwanted \emph{UP} check, keeping the authenticator busy and denying availability.
Since the attacker can detect when the busy state ends
(e.g., the user pressed the authenticator's button or 30 seconds have passed),
they can prolong the attack.

\textbf{AC$_{7}$: \aseven}.
In AC$_{7}$, the attacker invokes \cmd{GetInfo} to retrieve the authenticator's details.
Then, similar to CI$_{4}$, they use this information as a stepping stone to other attacks, tracks the user,
or checks whether the authenticator is vulnerable to implementation-specific attacks~\cite{yubico-cve}.
Not requiring \emph{UV} or \emph{UP}, the attacker can confound any API call into a \cmd{GetInfo} call.

\section{CTRAPS Vulnerabilities and Impact}\label{sec:vuln}

We present the root causes of the CTRAPS attacks
and discuss our attacks' impact on the FIDO2 ecosystem.

\subsection{Vulnerabilities}\label{subsec:vulns}

The CI and AC attacks are enabled by \emph{\vulns\ vulnerabilities} we discovered
in the CTAP specification. Seven of them are novel, whereas V2 was discussed in~\cite{peng2021display}.
Regardless, this is the first work exploiting V2 via AC.
Now, we will describe the vulnerabilities
and map them to the CI and AC attacks.

\textbf{V1: \vone}.
The CTAP client does not authenticate to the authenticator.
FIDO2 clients (and, by extension, CTAP clients) lack an identity,
preventing the authenticator from distinguishing an official client
developed by its manufacturer and a third-party client.
As a result, the authenticator trusts any connecting client,
including spoofed ones.

\textbf{V2: \vtwo}.
The authenticator does not provide visual feedback to the user
when invoking the CTAP Authenticator API.
As a result, the user is unable to verify whether the intended API
has been correctly called (or has been confounded instead),
or which API utilized the most recent \emph{UV} and \emph{UP} authorizations granted.

\textbf{V3: \vthree}.
Authenticators within the NFC range of a FIDO2 client automatically obtain \emph{UP}
without the user pressing a button on the authenticator.
Bypassing \emph{UP} means that \cmd{MakeCred},
\cmd{GetAssertion}, and \cmd{Reset}
are solely protected by \emph{UV},
or they now require no authorization at all.

\textbf{V4: \vfour}.
Destructive API calls, such as credential deletion (\cmd{CredMgmt})
or authenticator factory reset (\cmd{Reset}),
and non-destructive ones, like authentication (\cmd{GetAssertion})
are authorized by the same \emph{UV} PIN.
Whenever the user grants \emph{UV} for a non-destructive operation,
they unknowingly over-privilege the client,
enabling destructive operations as well.
For example, the user grants \emph{UV} to authenticate (non-destructive),
but an AC attacker exploits the over-privileged access
to instead factory reset the authenticator (destructive).

\textbf{V5: \vfive}.
Discoverable credentials contain \emph{static} and \emph{unique} CredId and UserId,
exploitable to reliably track users. These values can be obtained
without \emph{UV} or \emph{UP} via the \cmd{GetAssertion} API.
We note that the more credentials are stored in the authenticator,
the more this vulnerability is effective,
as each credential improves to the user's fingerprint.

\textbf{V6: \vsix}.
Despite being destructive, the \cmd{Reset} API \emph{only} requires \emph{UP},
which does not verify that the person operating the device is the owner.
Anyone nearby the authenticator can obtain \emph{UP} by pressing its button
or by being within NFC range.

\textbf{V7: \vseven}.
The \cmd{CredMgmt} API \emph{only} requires \emph{UV},
meaning that no user interaction is needed to delete discoverable credentials.
This allows multiple discoverable credentials to be deleted
without first alerting the user or asking for a confirmation.

\textbf{V8: \veight}.
The \cmd{Selection} API can be used to DoS an authenticator by continuously
prompting it for \emph{UP} checks.
\\
\\
\indent
Table~\ref{tab:mapping} maps the \vulns\ vulnerabilities (columns) to the \attacks\ \aname\
attacks (rows).
V1 is needed to perform all CI and AC attacks, as it allows an untrusted client
or a MitM attacker to connect to the authenticator without authenticating to it.
V2 provides stealthiness to AC attacks because, without visual feedback,
the user cannot confirm whether the API they are calling is being confounded or not.
Due to V3, CI$_{1}$ and CI$_{2}$ over NCF require zero clicks instead of one (\emph{UP}).
V3 also unlocks several new API confusion combinations,
such as \cmd{GetInfo} into \cmd{Reset}.

Moreover, V4 allows to perform the (destructive) CI$_{2}$, CI$_{3}$, AC$_{1}$, AC$_{2}$, and AC$_{5}$
attacks even when the user calls a non-destructive API, such as \cmd{Selection}.
V5 enables the usage of identifiers as persistent fingerprints,
resulting in two user tracking attacks (CI$_{2}$ and AC$_{3}$).
V6 allows for a zero-click factory reset attack (CI$_{1}$) over NFC.
V7 allows for a one-click credential deletion attack (AC$_{1}$).
V8 enables a persistent and reliable DoS attack on the authenticator (AC$_{6}$).

\begin{table}[tb]
\renewcommand{\arraystretch}{1.1}
\centering\small
\caption{Mapping the \vulns\ vulnerabilities (columns) to the four CI and seven AC attacks.}
\begin{tabular}{@{}ccccccccc@{}}
\toprule
        & \textbf{V1}    & \textbf{V2}    & \textbf{V3}    & \textbf{V4}    & \textbf{V5}    & \textbf{V6}   & \textbf{V7}   & \textbf{V8} \\
\midrule
CI$_{1}$ & \cmark\       & \cmark\        & \cmark\        & \xmark\        & \xmark\        & \cmark\       & \xmark\       & \xmark\        \\
CI$_{2}$ & \cmark\       & \cmark\        & \cmark\        & \xmark\        & \cmark\        & \xmark\       & \xmark\       & \xmark\        \\
CI$_{3}$ & \cmark\       & \cmark\        & \xmark\        & \xmark\        & \xmark\        & \xmark\       & \xmark\       & \xmark\        \\
CI$_{4}$ & \cmark\       & \cmark\        & \xmark\        & \xmark\        & \xmark\        & \xmark\       & \xmark\       & \xmark\        \\
\midrule
AC$_{1}$ & \cmark\       & \cmark\        & \xmark\        & \cmark\        & \xmark\        & \xmark\       & \cmark\       & \xmark\        \\
AC$_{2}$ & \cmark\       & \cmark\        & \cmark\        & \cmark\        & \xmark\        & \cmark\       & \xmark\       & \xmark\        \\
AC$_{3}$ & \cmark\       & \cmark\        & \cmark\        & \xmark\        & \cmark\        & \xmark\       & \xmark\       & \xmark\        \\
AC$_{4}$ & \cmark\       & \cmark\        & \cmark\        & \xmark\        & \xmark\        & \xmark\       & \xmark\       & \xmark\        \\
AC$_{5}$ & \cmark\       & \cmark\        & \xmark\        & \cmark\        & \xmark\        & \xmark\       & \xmark\       & \xmark\        \\
AC$_{6}$ & \cmark\       & \cmark\        & \xmark\        & \xmark\        & \xmark\        & \xmark\       & \xmark\       & \cmark\        \\
AC$_{7}$ & \cmark\       & \cmark\        & \xmark\        & \xmark\        & \xmark\        & \xmark\       & \xmark\       & \xmark\        \\
\bottomrule
\end{tabular}
\label{tab:mapping}
\end{table}

\subsection{Impact}\label{subsec:impact}

The \attacks\ \aname\ attacks break the security, privacy, and availability of the FIDO2 ecosystem,
with widespread and severe implications.
We support our claims with the experimental results presented in Section~\ref{sec:eval}.

Our attacks exploit \emph{protocol-level} CTAP vulnerabilities,
working regardless of the transport and the implementation details
of the authenticator, the client, and the relying party.
Hence, they threaten millions of authenticators in the wild,
their users, and the relying parties.
Being at the protocol-level, the root causes are challenging to fix, as most authenticators,
including YubiKeys, do not support firmware updates.

The CTRAPS attacks are \emph{practical} and \emph{low-cost},
requiring minimal equipment, such as a smartphone.
Their outcome is realistic and involves limited or no user interaction,
as shown in the video demonstrations available in the CTRAPS GitHub repository.
For instance, we lost access to our test Google and Apple ID accounts
because we could not pass 2FA after deleting our credentials with AC$_{1}$.

\subsection{Comparison with prior FIDO attacks}

Table~\ref{tab:rw-compare} compares our attacks with previous attacks on FIDO.
The CI attacks are the first client impersonation attacks targeting FIDO2 (CTAP2+),
while the AC attacks utilize API confusion, a novel attack strategy for FIDO2.
Prior research on CTAP evaluated only the \cmd{ClientPin} and \cmd{MakeCred} APIs.
Instead, our attacks target the entire Authenticator API, regardless of the CTAP transport,
covering a broader surface.

The CI attacks have low complexity, as they do not require a compromised client
or prior knowledge of user secrets (e.g., credential identifiers).
The AC attacks have a moderate complexity, as they require a MitM position.
Both CI and AC attacks have a high impact as they can, for example,
destroy credentials and track users.

The existing FIDO attacks with a high impact also require a strong attacker model,
such as physical access or malware installed on the user's browser or device.
In contrast, the CTRAPS attacks employ a \emph{weaker} attacker model,
i.e., client impersonation and MitM attacker, while still achieving high impact with low-to-mid complexity.

\section{Implementation}\label{sec:impl}

In this section, we present \toolkit, a novel toolkit that implements
the \aname\ attacks and enables experimentation with CTAP.
The toolkit has \emph{three} modules:
a CTAP testbed (Section~\ref{subsec:imp-virtual-test}),
four customizable CTAP clients (Section~\ref{subsec:imp-reader-impers}),
and an enhanced FIDO2 Wireshark dissector (Section~\ref{subsec:imp-dissect}).

The CTAP testbed and the Electron app CTAP client need
the user's authorization to connect and communicate with the authenticator.
Linux requires adding extra \texttt{udev} rules,
macOS asks to accept a notification on the screen,
and Windows needs admin privileges.
This limitation is expected, as it is also present in FIDO2 apps
released by authenticator manufacturers,
such as the Yubico Authenticator App and the Feitian Authenticator Tool.
Now, we will describe the implementation of each module and highlight their novelties.

\subsection{CTAP Testbed}\label{subsec:imp-virtual-test}

Our CTAP testbed includes a virtual WebAuthn relying party and a virtual
WebAuthn/CTAP client. The testbed can test real authenticators
without having to tamper with actual credentials and also launch the CTRAPS attacks.
Our relying party and client extend the Yubico open-source Python library for
FIDO2 called \texttt{python-fido2}~\cite{yubico-testbed}.

\textbf{Virtual relying party}.
The virtual relying party is implemented as a customizable WebAuthn server.
It includes standard relying party templates
and fast customization of the server's parameters.
For example, we implemented a template imitating a Microsoft relying party,
including its FIDO2 identifier (i.e., \emph{login.microsoft.com}).
The virtual relying party is useful to quickly test
real authenticators against CI and AC attacks.
For example, we can automatically register credentials
with different protection policies on the authenticator.

\textbf{Virtual client}.
The virtual client offers a convenient CTAP API,
offering low-level access to any CTAP message.
It can send CTAP commands in any order, or issue custom and malformed payloads.
It can be configured with different CTAP authorization requirements,
authentication challenges, and origins.

\subsection{CTAP Clients}\label{subsec:imp-reader-impers}

We developed four custom CTAP clients:
an Android app performing proximity CI over NFC,
an Android app performing remote CI over NFC,
a Proxmark3 script that executes proximity CI over NFC,
and an Electron app simulating a MitM attacker to test remote AC over USB.
We released in the CTRAPS GitHub repository five video demonstrations,
showing how to deploy the CTRAPS attacks on real authenticators using our clients.

\textbf{Android app for proximity CI over NFC}.
We implemented the proximity CI attacks using an Android app
which impersonates a FIDO2 client over NFC.
The app runs on a device owned by the attacker
and targets any authenticator that comes within the NFC range.
For example, it can perform CI$_{2}$ to leak identifiers and track the user.

\textbf{Android app for remote CI over NFC}.
We utilized an Android app to implement the remote CI attacks over NFC.
The app is installed on a device owned by the victim.
It spoofs a legitimate NFC app,
enticing the user to scan their authenticator (e.g., by asking for FIDO2 authentication).
The attacker can connect to the app and manage the CTAP connection with the authenticator.
The app does not need root privileges and asks at runtime
for the dangerous \texttt{\seqsplit{android.permission.NFC}},
required to gain access to the \texttt{\seqsplit{android.nfc}}~\cite{android-nfc} API.
However, this is not a concern, as the app is not trying to conceal its NFC capabilities.
The app also needs the standard install-time \texttt{\seqsplit{android.permission.INTERNET}}
to exfiltrate the data collected through CI$_{2}$ and CI$_{4}$.

\textbf{Proxmark3 for proximity CI over NFC}.
We implemented the proximity CI attacks using the Proxmark3~\cite{nfc-proxmark},
an open-source and programmable development kit for NFC (RFID).
We wrote a Lua script using the Proxmark3 ISO14443 Type A module (i.e., \texttt{read14a})
to communicate to the authenticator via CTAP-compliant APDUs.
By equipping the Proxmark3 with a long-range high-frequency antenna,
we were able to extend its reach.
The long-range antenna has an indicative range of 100 to 120 millimeters,
as opposed to the 40 to 85 millimeters of the built-in antenna.

\textbf{Electron app to simulate AC over USB}.
We developed an Electron app that simulates a MitM attacker.
The app uses the \emph{node-hid} module to access the USB HID traffic,
gaining a MitM position between a FIDO client and an authenticator communicating over USB.
The app scans for local HID devices and identifies the authenticators from their
properties (e.g., the product and manufacturer fields).
Then, it connects and sends binary data over USB to the authenticator.
The Electron app is compatible with Windows, macOS, and Linux.

\subsection{FIDO2 Wireshark Dissector}\label{subsec:imp-dissect}

We extended an unofficial Wireshark FIDO2 dissector found in~\cite{dissect-fido}.
We add valuable features, such as support for the \cmd{CredMgmt} API.
We include parsers for \texttt{WAITING} and \texttt{PROCESSING} keepalive status codes
that identify when authenticators are unavailable.
We parse the authenticator's capabilities in the \texttt{CTAPHID\_INIT} message,
which are useful for testing AC$_{7}$.
We provide an improved way to display CTAP data
when dissecting CTAPHID (USB) and ISO7816/ISO14443 (NFC).
Finally, we add missing vendor and product identifiers to the dissector tables.
The FIDO2 dissector is included in our toolkit as a Lua script (i.e., \texttt{fido2-dissectors.lua}).

\section{Evaluation}\label{sec:eval}

We evaluated our \attacks\ attacks against \emph{\auths} popular and recent
authenticators from Yubico, Feitian, SoloKeys, and Google.
We also tested \emph{\rps} widely used relying parties,
including Microsoft, Apple, GitHub, and Facebook.
Next, we will present our evaluation setup and results.

\subsection{Setup}\label{subsec:eval-setup}

\textbf{Authenticators}.
We evaluate \emph{\auths} popular FIDO2 authenticators.
Table~\ref{tab:setup-auths} shows their technical details.
The YubiKey 5 NFC, YubiKey 5 NFC FIPS, and Feitian NFC K9
are closed-source and do not support firmware updates.
The Solo V1, Solo V2 Hacker, and Open Security Key (OpenSK) have an
open-source firmware (OSF), that we updated to their latest version.
The authenticators support USB and NFC,
except for OpenSK which has an NFC module but supports only USB.
The Solo V1 requires a button press to activate NFC.
Unfortunately, we could not find any FIDO2 authenticator supporting BLE.

The authenticators store a maximum of 25 (Yubico), 50 (Feitian and SoloKeys),
or 150 (OpenSK) discoverable credentials.
The YubiKey 5 FIPS is FIPS140-2 compliant, and,
as such, it should provide high security guarantees.
We ran OpenSK on an NRF52840 dongle,
but any board supporting OpenSK would have worked.

\begin{table}[tb]
  \caption{Details about the \auths\ authenticators we attack. All
  authenticators support USB and NFC, except OpenSK, which only supports USB. FVer:
  firmware version, OSF: open-source firmware, DCr: discoverable credentials.}
  \renewcommand{\arraystretch}{1.1}
  \centering\small
    \begin{tabular}{@{}llrrrr@{}}
    \toprule
    \textbf{Authenticator} & \textbf{Manuf} & \textbf{Year} & \textbf{FVer} & \textbf{OSF}  & \textbf{DCr} \\
    \midrule
    YubiKey 5              & Yubico         & 2018          & 5.2.7         & No            & 25             \\
    YubiKey 5 FIPS         & Yubico         & 2021          & 5.4.3         & No            & 25             \\
    Feitian K9             & Feitian        & 2016          & 3.3.01        & No            & 50             \\
    Solo V1                & SoloKeys       & 2018          & 4.1.5         & Yes           & 50             \\
    Solo V2 Hacker         & SoloKeys       & 2021          & 2.964         & Yes           & 50             \\
    OpenSK                 & Google         & 2023          & 2.1           & Yes           & 150            \\
    \bottomrule
  \end{tabular}
\label{tab:setup-auths}
\end{table}

\begin{table*}[tb]
    \caption{CI and AC attacks on \auths\ authenticators.
        The first column lists the authenticators' names.
        The remaining columns report our four CI and seven AC attacks on CTAP.
        \cmark: attack is effective on the authenticator,
    \textbf{n/a}: not applicable as the authenticator does not implement the \texttt{Selection} API.}
    \renewcommand{\arraystretch}{1.1}
    \centering\small
    \begin{tabular}{@{}lccccccccccc@{}}
        \toprule
        \textbf{Authenticator} & \textbf{CI$_{1}$} & \textbf{CI$_{2}$} & \textbf{CI$_{3}$} & \textbf{CI$_{4}$}
                         & \textbf{AC$_{1}$} & \textbf{AC$_{2}$} & \textbf{AC$_{3}$} & \textbf{AC$_{4}$} & \textbf{AC$_{5}$} & \textbf{AC$_{6}$} & \textbf{AC$_{7}$} \\
                         \midrule
        YubiKey 5                          & \cmark\          & \cmark\           & \cmark\          & \cmark\
                                           & \cmark\     & \cmark\     & \cmark\     & \cmark\     & \cmark\     & n/a         & \cmark\     \\
        YubiKey 5 FIPS                 & \cmark\          & \cmark\           & \cmark\          & \cmark\
                                       & \cmark\     & \cmark\     & \cmark\     & \cmark\     & \cmark\     & n/a         & \cmark\     \\
        Feitian K9               & \cmark\          & \cmark\           & \cmark\                      & \cmark\
                                 & \cmark\     & \cmark\     & \cmark\     & \cmark\     & \cmark\     & n/a         & \cmark\     \\
        Solo V1                   & \cmark\          & \cmark\           & \cmark\                       & \cmark\
                                  & \cmark\     & \cmark\     & \cmark\     & \cmark\     & \cmark\     & n/a         & \cmark\     \\
        Solo V2 Hacker                & \cmark\          & \cmark\           & \cmark\            & \cmark\
                                      & \cmark\     & \cmark\     & \cmark\     & \cmark\     & \cmark\     & \cmark\     & \cmark\     \\
        OpenSK                  & \cmark\          & \cmark\           & \cmark\                        & \cmark\
                                & \cmark\     & \cmark\     & \cmark\     & \cmark\     & \cmark\     & \cmark\     & \cmark\     \\
                                \bottomrule
    \end{tabular}
    \\[0.1cm]
    \textbf{CI$_{1}$}: \atwo, \textbf{CI$_{2}$}: \athree, \textbf{CI$_{3}$}: \afive, \textbf{CI$_{4}$}: \aseven,
    \textbf{AC$_{1}$}: \aone, \textbf{AC$_{2}$}: \atwo, \textbf{AC$_{3}$}: \athree,
    \textbf{AC$_{4}$}: \afour, \textbf{AC$_{5}$}: \afive, \textbf{AC$_{6}$:} \asix, \textbf{AC$_{7}$}: \aseven.
    \label{tab:results-auths}
\end{table*}

\textbf{Relying parties}.
Our list of relying parties covers pervasive and heterogeneous online services,
including software as a service, social, gaming,
cryptographic signing, authentication, and cloud storage.
We registered our authenticators with \emph{\rps} FIDO2 relying parties: Adobe, Apple,
DocuSign, Facebook, GitHub, Hancock, Microsoft, NVidia, Synology, and Vault Vision.
Some of them offer Single Sign-On (SSO), enabling access to multiple services.
For example, a single set of FIDO2 credentials
can log into Microsoft, OneDrive, Outlook, and Minecraft.
As a consequence, erasing a single credential has a widespread effect on multiple online services.

\textbf{\aname\ toolkit}.
We evaluated the CI and AC attacks using the tools included in the CTRAPS toolkit.
We installed our two Android apps, found in the CTAP clients module,
on a Google Pixel 2 (OS: Android 11),
a RealMe 11 Pro (OS: Android 14), and a Xiaomi Redmi Plus 5 (OS: Android 8.1).
We used a Proxmark3 RDV4 with a long-range high-frequency antenna
to deploy the proximity CI attack with an extended NFC range.
We tested our Electron app on a Dell Inspiron 15 3502 laptop (OSes: Ubuntu 22.04.3 LTS
and Windows 11 Home) and on a MacBook Pro M1 (OS: macOS Ventura 13.4).

\subsection{Authenticators Results}\label{subsec:eval-au}

Table~\ref{tab:results-auths} shows the evaluation results for
the CI and AC attacks on six FIDO2 authenticators.
All six of them were vulnerable to the CTRAPS attacks,
even the FIPS-compliant YubiKey.
As expected, since we attack CTAP at the protocol level,
the attacks are effective regardless of the CTAP transport (i.e., USB or NFC),
or the authenticator's software and hardware.
However, $AC_6$ does not apply to the four authenticators
which do not support the \cmd{Selection} API.

We also found a \cmd{CredMgmt} implementation vulnerability
on the YubiKey 5 and YubiKey 5 FIPS,
which improperly handles the authenticator's state for \cmd{CredMgmt}.
They allow the client to call \cmd{CredMgmt(EnumRpsGetNextRp)}
without invoking \cmd{CredMgmt(EnumRpsBegin)} first, which is an illegal state.
We exploit this flaw to achieve a zero-click \emph{leak of relying party names}.
Our attack calls \cmd{CredMgmt(EnumRpsGetNextRp)}
to reveal the names of all the relying parties, stored on the authenticator, except one.
This attack bypasses \emph{UV} and works regardless of the \emph{CredProtect} policy.
We reported it to Yubico, which assigned it CVE-2024-35311
and addressed it in their latest firmware.
However, since YubiKeys do not support firmware updates,
this fix is only available to newer authenticators, leaving older ones vulnerable.

The CI and AC attacks over NFC have a maximum range of two centimeters on a smartphone.
The Proxmark3 built-in antenna also achieved the same range,
which we could extend to six and a half centimeters by attaching a long-range antenna.
Prior work demonstrated that, with specialized equipment,
the NFC range can be extended up to 50 centimeters~\cite{kfir2005range}.

We also tested \emph{combinations} of CI and AC attacks,
to develop more advanced variants.
For example, we found multiple ways to enhance our user tracking attacks (CI$_{2}$ and AC$_{3}$).
The attacker can refine the user's fingerprint using AC$_{7}$
or register new credentials, with metadata of their choice, on the authenticator using AC$_{4}$.

\subsection{Relying Parties Results}\label{subsec:eval-rp}

As shown in Table~\ref{tab:results-rps}, we tested \discrps\ relying parties supporting discoverable credentials
and \ndiscrps\ employing non-discoverable credentials.
Our evaluation includes relying parties because our attacks affect them,
even though we do not utilize WebAuthn.
For example, AC$_{1}$ deletes discoverable credentials,
causing the user to lose access to their online account.
Although relying parties using non-discoverable credentials are not vulnerable to AC$_{1}$ and AC$_{4}$,
they remain open to our factory reset, user tracking, and DoS attacks.

\begin{table*}[tb]
  \caption{\aname\ attacks on \rps\ relying parties.
  The first and second columns list the relying parties' names and identifiers.
  The third column highlights whether they register discoverable (Disc, DiscWeak)
  or non-discoverable (NonDisc) credentials. We indicate with DiscWeak a relying party
  using the default and weak \emph{CredProtect=UVOptional} policy.
  Columns four, five, and six specify the effect of each attack.
  n/a: the attack is not applicable because the relying party does not support discoverable credentials.}
  \renewcommand{\arraystretch}{1.1}
  \centering\small
    \begin{tabular}{@{}llllll@{}}
    \toprule
    \textbf{Rp} & \textbf{RpId}                                 & \textbf{Cred}
                                      & \textbf{Delete Creds}                 & \textbf{Track User}                 & \textbf{DoS Authenticator} \\
    \midrule
    Adobe                     & \texttt{adobe.com}                      & Disc
    				& CI$_{1}$, AC$_{1}$, AC$_{2}$                                 & CI$_{2}$, AC$_{3}$                                    & CI$_{3}$, AC$_{4}$, AC$_{5}$, AC$_{6}$ \\
    Apple                      & \texttt{apple.com}                        & DiscWeak
    				& CI$_{1}$, AC$_{1}$, AC$_{2}$                                 & CI$_{2}$, AC$_{3}$                                    & CI$_{3}$, AC$_{4}$, AC$_{5}$, AC$_{6}$ \\
    DocuSign               & \texttt{account.docusign.com} & NonDisc
    				& CI$_{1}$, AC$_{2}$                                           & n/a                                                & CI$_{3}$, AC$_{5}$, AC$_{6}$ \\
    Facebook               & \texttt{facebook.com}                & NonDisc
    				& CI$_{1}$, AC$_{2}$                                          & n/a                                                 & CI$_{3}$, AC$_{5}$, AC$_{6}$ \\
    GitHub                    & \texttt{github.com}                     & Disc
    				& CI$_{1}$, AC$_{1}$, AC$_{2}$                                & CI$_{2}$, AC$_{3}$                                    & CI$_{3}$, AC$_{4}$, AC$_{5}$, AC$_{6}$ \\
    Hancock                 & \texttt{hancock.ink}                     & Disc
    				& CI$_{1}$, AC$_{1}$, AC$_{2}$                                 & CI$_{2}$, AC$_{3}$                                    & CI$_{3}$, AC$_{4}$, AC$_{5}$, AC$_{6}$ \\
    Microsoft              & \texttt{login.microsoft.com}     & DiscWeak
    				& CI$_{1}$, AC$_{1}$, AC$_{2}$                                & CI$_{2}$, AC$_{3}$                                    & CI$_{3}$, AC$_{4}$, AC$_{5}$, AC$_{6}$ \\
    NVidia                    & \texttt{login.nvgs.nvidia.com}  & Disc
    				& CI$_{1}$, AC$_{1}$, AC$_{2}$                                 & CI$_{2}$, AC$_{3}$                                    & CI$_{3}$, AC$_{4}$, AC$_{5}$, AC$_{6}$ \\
    Synology               & \texttt{account.synology.com} & Disc
    				& CI$_{1}$, AC$_{1}$, AC$_{2}$                               & CI$_{2}$, AC$_{3}$                                    & CI$_{3}$, AC$_{4}$, AC$_{5}$, AC$_{6}$ \\
    Vault Vision         & \texttt{auth.vaultvision.com}    & Disc
    				& CI$_{1}$, AC$_{1}$, AC$_{2}$                               & CI$_{2}$, AC$_{3}$                                    & CI$_{3}$, AC$_{4}$, AC$_{5}$, AC$_{6}$ \\
    \bottomrule
  \end{tabular}
\label{tab:results-rps}
\end{table*}

CI$_{1}$, AC$_{1}$, and AC$_{2}$ block web authentication to the relying party by deleting the user's FIDO2 credentials.
CI$_{2}$ and AC$_{3}$ utilize the user identifiers generated by the relying party to track users.
CI$_{3}$, AC$_{4}$, AC$_{5}$, and AC$_{6}$ prevent relying parties from communicating with the authenticator.
Among the relying parties supporting discoverable credentials, we found that only Microsoft and Apple
employ the weak \emph{CredProtect=UVOptional} policy.
This policy allows to bypass \emph{UV} when accessing credentials.
As a result, an attacker can deploy a zero-click variant of CI$_{2}$ and AC$_{3}$ to track users through their Microsoft and Apple credentials.

\section{Discussion}\label{sec:discuss}

We discuss the countermeasures to fix the CTRAPS attacks
and the issues we found in the FIDO reference threat model.

\subsection{Countermeasures}\label{sec:counters}

We discuss \emph{\counters} backward-compliant countermeasures
fixing the \attacks\ CTRAPS attacks and their associated \vulns\ vulnerabilities.
Each countermeasure addresses a specific vulnerability (e.g., C1 fixes V1)
and helps reduce the CTAP attack surface.
Although we have not implemented these countermeasures,
we designed them to be implementable as amendments to the FIDO2 standard
or as FIDO2 extensions.
Next, we will describe each countermeasure.

\textbf{C1: \cone}.
We address V1 by recommending that the FIDO Alliance provides a list of trusted CTAP clients.
The FIDO ecosystem offers several certifications,
including the FIDO Functional Certification~\cite{fidoall-functional-cert}
which attests to the \emph{interoperability} of clients, servers, and authenticators.
We suggest extending this certification to also cover the \emph{trustworthiness} of CTAP clients.
For instance, FIDO could implement a Software Bill Of Materials (SBOM) solution
to monitor trusted CTAP clients and their vulnerabilities~\cite{wu2023sbom}.

\textbf{C2: \ctwo}.
We address V2 by requiring the authenticator to provide the user
with visual feedback regarding the API that was called.
For instance, the authenticator's LED could blink \emph{once} for non-destructive API calls
and \emph{twice} for destructive ones.
The CTAP \emph{wink} command, which blinks the LED, must be disabled during this visual feedback step.

\textbf{C3: \cthree}.
We address V3 by requiring user interaction during \emph{UP} checks over NFC.
For example, the user could press a button on the authenticator to grant \emph{UP} over NFC,
similar to \emph{UP} checks over USB.

\textbf{C4: \cfour}.
We address V4 by introducing a dedicated PIN to authorize
destructive API calls (e.g., \cmd{CredMgmt} and \cmd{Reset}) and by repurposing
the current PIN to authorize non-destructive API calls (e.g., \cmd{Selection} and \cmd{GetInfo}).
The new PIN should have the same or stricter
requirements as the non-destructive PIN (i.e., four to sixty-three Unicode characters~\cite{ctap21-standard}).

\textbf{C5: \cfive}.
We address V5 by implementing dynamic CredId and UserId
and mandating \emph{\seqsplit{CredProtect=UVRequired}}.
CredId and UserId should rotate after a set amount of logins (e.g., every ten logins)
or a time interval (e.g., once per month).
Hence, we raise the bar for user profiling and tracking attacks on authenticators.
Currently, the user can indirectly change a CredId by calling \cmd{MakeCred}
to generate a new credential for their account, replacing the old one.
However, the user cannot change the UserId, which is determined by the relying party
and, based on our experience, remains fixed to the user account.

\textbf{C6: \csix}.
We address V6 by requiring \emph{UV} to call \cmd{Reset}.
Hence, the user must authorize a factory reset by entering a valid PIN.

\textbf{C7: \cseven}.
We address V7 by requiring \emph{UP} to call \cmd{CredMgmt}.
Hence, the user must authorize credential deletion one by one,
to avoid deleting multiple credentials with a single API call.

\textbf{C8: \ceight}.
We address V8 by enforcing temporal rate limiting on \cmd{Selection}
to a maximum of three calls within two minutes.
We are not expecting issues with our rate limiting,
akin to the limiting already existing for \cmd{ClientPin(GetPinToken)},
as a client typically calls \cmd{Selection} once per session.

\textbf{Usability of the countermeasures}.
The deployment of C1 and C8 does not affect usability.
C2 requires the user to notice the authenticator's visual feedback.
Implementing C3 introduces an additional \emph{UP} check each time
the client connects to the authenticator over NFC, which is costly.
C4 forces the user to remember a second PIN.
C5 introduces one \emph{UV} and \emph{UP} check each time
the credential and user identifiers are being rotated out, e.g., once per month.
The implementation of C6 would require PIN verification for every call to \cmd{Reset}.
C7 would add a button press each time the \cmd{CredMgmt} API is invoked.

\textbf{Authenticator with a display}.
We do not consider adding a display to a roaming authenticator
in the list of countermeasures as is not backward-compliant and would
require to recall all vulnerable authenticators without a display.
Moreover, for newer authenticators, it entails significant hardware and software modifications,
such as adding a secure display, a secure display controller firmware, and a battery,
that would introduce usability, performance, and cost issues.

\begin{table*}[tb]
  \caption{Comparing prior attacks on FIDO with the CTRAPS attacks.
  We assign each attack a complexity and an impact.
  For example, the complexity for a MitM is Mid,
  whereas we consider spoofing a client as Low complexity.
  Similarly, hijacking a session has a Mid impact,
  while permanently destroying credentials carries a High impact.}
  \renewcommand{\arraystretch}{1.1}
  \centering\small
  \begin{tabular}{@{}lllllclll@{}}
    \toprule
	\textbf{Attack} & \textbf{Class} & \textbf{Protocol} & \textbf{Transp} & \textbf{Surface} & \textbf{Impl}
		 & \textbf{Reqs} & \textbf{Complex} & \textbf{Impact} \\
    \midrule
    		CTAP MitM~\cite{guan2022formal} & DH MitM & CTAP2.0 & All & ClientPin & \xmark
    			& MitM & Mid & Mid \\
		Privacy leak~\cite{guan2022formal} & Eavesdropping & CTAP2.0 & All & MakeCred & \xmark
    			& n/a & Low & Low \\
    		Auth rebind~\cite{guan2022formal} & Auth rebind & WebAuthn & All & Creds.create & \xmark
    			& n/a & High & High \\
    		Parallel session~\cite{guan2022formal} & Session hijack & WebAuthn & All & Creds.get & \xmark
    			& n/a & Mid & Mid \\

		ECDSA extract~\cite{titanninjaphysical} & Side channel & n/a & n/a & NXP A7005 & \xmark
			& Phy access & High & High \\
    		Titan sign in~\cite{google-titanble} & Relay & U2F & BLE & Google acc & \cmark
    			& Proximity & Mid & Mid \\
		Evil maid~\cite{lomne-hwio22} & Phy access & n/a & n/a & Auth TEE & \xmark
			& Phy access & High & High \\

		Auth MitM~\cite{barbosa2023rogue} & DH MitM & CTAP2.0/2.1 & USB & ClientPin & \cmark
    			& Mal browser & Mid & Mid \\
		Web MitM~\cite{barbosa2023rogue} & Session hijack & WebAuthn & USB & Creds.get & \cmark
    			& Mal browser & Mid & Mid \\
    		Rogue key~\cite{barbosa2023rogue} & Auth rebind & WebAuthn & USB & Creds.create & \cmark
    			& Mal browser & Mid & High \\

		FIDOLA~\cite{mahdad2024overlays} & Session hijack & WebAuthn & USB & Creds.get & \cmark
    			& Malware & High & Mid \\

    \textbf{CTRAPS CI} & Impersonation & CTAP2.0/2.1/2.2 & All & Auth API & \cmark
    			& Proximity & Low & High \\
		\textbf{CTRAPS AC} & API confusion & CTAP2.0/2.1/2.2 & All & Auth API & \cmark
    			& MitM & Mid & High \\
    \bottomrule
  \end{tabular}
\label{tab:rw-compare}
\end{table*}

\subsection{FIDO Reference Threat Model Issues}\label{app:frtm}

The FIDO Alliance released a reference threat model~\cite{fidoall-secref} outlining security assumptions,
goals, and threats against clients, authenticators, and relying parties.
Although non-normative, it is the only official source detailing the FIDO threat model.
After studying it and working on the CTRAPS attacks,
we identified \emph{three issues} (IS1, IS2, and IS3) with the FIDO reference threat model:

\textbf{IS1: Unclear security boundaries}.
The threat model presents six broad security assumptions,
but breaks them when discussing threats.
For example, SA-4 states that the user device and applications
involved in a FIDO2 operation act as trustworthy agents of the user.
This implies that the client (e.g., browser or mobile app) must be inherently trusted.
However, at the same time, the threat model includes threats that violate SA-4,
such as \emph{T-1.2.1: FIDO client corruption}.
This leads to unclear security boundaries,
making it difficult to differentiate trusted components
from ones that could be compromised.

\textbf{IS2: Missing proximity threats}.
Although FIDO supports proximity transports like NFC and BLE,
its threat model groups proximity-based threats together with physical access ones.
However, these threats differ in key aspects.
For example, proximity threats have a range.
Consequently, our proximity CI and AC attacks do not fit within this threat model.

\textbf{IS3: Security goals are narrow}.
The security goals of the threat model are based
on~\cite{fidoall-authschemes} (2006) and~\cite{bonneau2012quest} (2012).
These two research papers outlined the security goals of an ideal authentication scheme,
focusing on password-based schemes and web authentication.
As a result, the security goals are too narrow to capture the complexities of the FIDO ecosystem.
For example, there are no security goals for the Authenticator API, that could address the AC attacks,
or discoverable credentials, that are relevant to AC$_{1}$, CI$_{1}$, and AC$_{2}$.

\section{Related Work}\label{sec:related}

We present related work on FIDO, covering existing attacks, formal analysis,
FIDO extensions and enhancements, usability studies, and surveys.

\textbf{Attacks on FIDO(2)}.
Researchers found attacks on older FIDO versions (UAF, U2F), such as authenticator rebinding,
parallel sessions, and multi-user attacks~\cite{hu2016security,li2020authenticator},
USB HID man-in-the-middle attacks~\cite{bui2018man}, BLE pairing~\cite{google-titanble},
relying party public key substitution~\cite{scott2021fido},
bypassing push-based 2FA~\cite{jubur2021bypassing}, real-time phishing~\cite{ulqinaku2021real},
and side channel attacks~\cite{roche2021side,kepkowski2022not}.
FIDO2 was also found vulnerable to deception~\cite{mahdad2024overlays}, misbinding~\cite{yadav2023security},
physical~\cite{titanninjaphysical,lomne-hwio22},
and rogue key or impersonation attacks~\cite{kuchhal2023evaluating,barbosa2023rogue}.
Moreover, researchers found issues on lower layers trusted by FIDO2,
including an IV reuse on the Samsung Keystore~\cite{shakevsky2022trust}.
No prior attack investigated \emph{client impersonation} or \emph{API confusion} on CTAP,
including its \emph{latest} version.

\textbf{Formal analysis}.
The formal analysis and verification community extensively researched FIDO. The
community formally verified FIDO's Universal Authentication Framework
(UAF)~\cite{pereira2018formal,feng2021formal},
FIDO2 (including its privacy, revocation, attestation, and post-quantum
crypto)~\cite{barbosa2021provable,bindel2023fido2,hanzlik2023token,bindel2023attest}.
Yubico proposed a key recovery mechanism based on a backup authenticator
that was proven secure using the asynchronous remote key generation (ARKG)
primitive~\cite{frymann2020asynchronous}.
Existing research on formal analysis is \emph{not} covering our CI and AC attacks.

\textbf{Extensions}.
FIDO supports extensions to add optional features in a backward-compliant way.
For instance, FeIDO~\cite{schwarz2022feido} proposes an extension to recover a
FIDO2 credential using an electronic identifier. Extensions are not secure by
default, and researchers proposed a fix to protect them against MitM
attacks~\cite{buttner2022protecting}.
We suggest to \emph{update} the CTAP specification rather than implementing our
countermeasures as FIDO extensions that would be optional and insecure by design.

\textbf{Enhancements}.
Researchers proposed (cryptographic) enhancements to FIDO protocols.
In~\cite{ghinea2023hybrid}, the authors present a hybrid post-quantum signature
scheme for FIDO2 and tested it using OpenSK~\cite{google-opensk} (which we
exploit in this work). In~\cite{hanzlik2023revoke}, the authors propose
a global key revocation procedure for WebAuthn that revokes credentials
without communicating to each individual relying party WebAuthn server.
True2F~\cite{dauterman2019true2f} presented a backdoor-resistant FIDO U2F
design, protecting the authenticator from a malicious browser by requiring the authenticator
interaction during every authentication, and from fingerprinting by rate limiting
credential registration.
Proposed enhancements are \emph{not} addressing our attacks,
which are effective \emph{regardless of} the FIDO2 cryptographic primitives.

\textbf{Usability}.
Researchers performed extensive usability studies on FIDO
U2F~\cite{colnago2018s,das2018johnny,ciolino2019two,lassak2021s}, FIDO2
roaming authenticators~\cite{farke2020you,owens2021user},
passkeys~\cite{kepkowski2023challenges}, and cross-site 2FA~\cite{lyastani2023systematic}.
Our paper is \emph{orthogonal} to usability studies.

\textbf{Surveys}.
There are several FIDO survey papers.
In~\cite{angelogianni2021many}
the authors describe the evolution of FIDO protocols, security
requirements, and adoption factors.
In~\cite{lyastani2020fido2}, the authors surveyed the adoption of passwordless
authentication among a large user base, considering users' perceptions,
acceptance, and concern with single-factor authentication without passwords.
Our paper is \emph{orthogonal} to surveys.

\section{Conclusion}\label{sec:conclusion}

No prior work assessed the CTAP Authenticator API, a critical surface exposed by a
client to an authenticator to manage, create, and delete credentials.
We address this gap by presenting the first security and privacy evaluation
of the CTAP Authenticator API. We uncover two classes of protocol-level
attacks that abuse it. The CI attacks spoof a CTAP client to a target authenticator.
The AC attacks leverage a MitM position to change CTAP API calls made by the user
to an API desired by the attacker while stealing their authorizations.
They utilize API confusion, a novel attack strategy within FIDO2.

We uncover \attacks\ CI and AC attacks, impacting millions of FIDO2 users.
They can be deployed by a proximity-based or a remote attacker.
For example, they delete FIDO2 credentials and master keys (security breach)
and track users through their credentials (privacy breach).
Our attacks are effective on the entire FIDO2 ecosystem as they target
\vulns\ vulnerabilities we discovered in the CTAP specification.
These flaws include the lack of CTAP client authentication and improper API authorizations.
The \aname\ attacks are low-cost, as they do not require specialized equipment,
and stealthy, as they do not trigger unexpected user interactions.

We develop the \toolkit\ toolkit to test our attacks with a cheap setup.
It includes a CTAP testbed with a virtual relying party and a virtual client,
four CTAP clients that deploy our attacks (e.g., Android apps and Proxmark3 scripts),
and an enhanced Wireshark dissector for CTAP.
We successfully exploit \auths\ authenticators and \rps\ relying parties
from leading FIDO2 players such as Yubico, Feitian, Google, Microsoft, and Apple.
We design \counters\ legacy-compliant countermeasures
to fix our attacks and their root causes.

We share \emph{three lessons} we learned about
FIDO2 \emph{credential storage} and \emph{passwordless-ness},
which are valuable for the current transition from single-factor
authentication to 2FA and passkeys~\cite{gh-2fa,google-passkeys}:
(i) Being stored on the authenticator, FIDO2 discoverable credentials
are protected from third-party data breaches.
However, this introduces new attacks that work
exclusively on discoverable credentials (i.e.,  CI$_{2}$, AC$_{1}$, AC$_{3}$, and AC$_{4}$).
(ii) FIDO2 users cannot prevent attacks targeting discoverable credentials,
as they cannot choose the type of credentials they register and their protection policies,
decided by the relying party and the client instead.
(iii) The FIDO2 core message is to steer away from passwords because they are vulnerable to phishing.
However, digging deeper, we realized that FIDO2 still relies on phishable mechanisms, even
for passwordless authentication. For instance, a passwordless credential is protected by an
alphanumeric PIN (i.e., a phishable sequence the user must remember).

\section*{Acknowledgment}

Work funded by the European Union under grant agreement no. 101070008 (ORSHIN
project). Views and opinions expressed are however those of the author(s)
only and do not necessarily reflect those of the European Union. Neither the
European Union nor the granting authority can be held responsible for them.
Moreover, it has been partially supported by the French National Research
Agency under the France 2030 label (NF-HiSec ANR-22-PEFT-0009) and the
Apricot/ENCOPIA ANR MESRI-BMBF project (ANR-20-CYAL-0001).

\bibliographystyle{plain}
\bibliography{biblio}

\appendices

\section{Data Availability}

All experiments in this study were conducted ethically and solely on authenticators and accounts under our control,
with no involvement of third-party personal data.
To support reproducibility and advance open science,
we are releasing our artifacts, including the \aname\ toolkit.
Our toolkit is securely hosted in a repository at \url{https://github.com/Skiti/CTrAPs}.
We already responsibly disclosed our findings to all affected parties,
including the FIDO Alliance, and we respected their timeline.

\end{document}